\documentclass[journal]{journal}

\usepackage[outdir=./]{epstopdf}
\usepackage[pdftex]{graphicx}
\DeclareGraphicsExtensions{.pdf}
\ifCLASSINFOpdf
\else
\fi
\usepackage[cmex10]{amsmath}
\usepackage{amsthm}
\usepackage{amssymb}
\begin{document}
%
\title{Glushkov's construction for functional subsequential transducers}
%
%
%

\author{Aleksander Mendoza-Drosik}

\newtheorem{theorem}{Theorem}
\newtheorem{definition}{Definition}
%
%

\markboth{Journal of \LaTeX\ Class Files,~Vol.~6, No.~1, January~2007}%
{Shell \MakeLowercase{\textit{et al.}}: Bare Demo of IEEEtran.cls for Journals}
%



\maketitle
\thispagestyle{empty}

\begin{abstract}
 Glushkov's construction has many interesting properties and they become even more evident when applied to transducers. This article strives to show the vast range of possible extensions and optimisations for this algorithm.  Special flavour of regular expressions is introduced, which can be efficiently converted to $\epsilon$-free functional subsequential weighted finite state transducers. Produced automata are very compact, as they contain only one state for each symbol (from input alphabet) of original expression and only one transition for each range of symbols, no matter how large. Such compactified ranges of transitions allow for efficient binary search lookup during automaton evaluation. All the methods and algorithms presented here were used to implement open-source compiler of regular expressions for multitape transducers. 
\end{abstract}


%
\IEEEpeerreviewmaketitle

\section{Introduction}
%
%
%
%
\IEEEPARstart{T}{here}
 are not many open source solutions available for working with transducers. The most significant and widely used library is OpenFst. Their approach is based on theory of weighted automata\cite{MOHRI3}\cite{DROSTE}\cite{DROSTE2}. Here we propose an alternative approach founded on lexicographic transducers \cite{MendozaDrosik2020MultitapeAA} and Glushkov's algorithm \cite{GLUSHKOV}.

Let $W$ be some set of weight symbols. The free monoid $W^*$ will be out set of weight strings. We assume there is some lexicographic order defined as
\[
b_1w_1 > b_2w_2 \iff w_1 > w_2 \mbox{ or }( w_1=w_2\mbox{ and } b_1 > b_2) \\
\]
where $w_1,w_1\in W$ and $b_1,b_2\in W^*$.  The order is defined only on strings of equal lengths.
Let $\Sigma$ be the input alphabet, $\Sigma^*$ is the monoid of input strings and $D$ is the monoid of output strings.  Lexicographic transducer is defined as tuple $(Q,I,W,\Sigma,D,\delta,\tau)$ where $Q$ is some finite set of states, $I$ is the set of initial states, $\tau$ is a state output (partial) function $Q\rightarrow D \times W$ and lastly $\delta$ represents transitions of the form $\delta \subset Q \times W \times \Sigma \times D \times Q$.

Thanks to $\tau$, such machines are subsequential \cite{MOHRI}\cite{MOHRI2}\cite{HANSAN}\cite{de_la_higuera}. As an example consider the simple transducer from figure \ref{transducer}. The states $q_0$, $q_1$ and $q_2$ have no output, which can be denoted with $\tau(q_0)=\emptyset$. The only set which does have output is $q_3$. Every time automaton finishes reading input string and reaches $q_3$, it will append $d_0$ to its output and then accept. For instance, on input $\sigma_1\sigma_2$ it will first read $\sigma_1$, produce output $d_0d_4$ and go to state $q_1$, then read $\sigma_2$ and append output $d_3$, go to state $q_3$, finally reaching end of input, appending $d_0$ and accepting. The total output would be $d_0d_4d_3d_0$. Note that the automaton is nondeterministic, as it could take alternative route passing through $q_2$ and producing $d_3d_0$. In such scenarios weights are used to disambiguate output. The first route produces weight string $w_2w_3w_1$, while the second produces $w_3w_2w_1$. According to our definition of lexicographic order we have $w_2w_3w_1 > w_3w_2w_1$ (assuming that $w_3>w_2$). Throughout this article we will consider smaller weights to be "better". Hence the automaton should choose $d_3d_0$ as the definitive output for input $\sigma_1\sigma_2$. There might be situations in which two different routes have the exact same (equally highest) weight while also producing different outputs. In such cases, the automaton is ambiguous and produces multiple outputs for one input.


\section{Expressive power}

There are some remarks to be made about lexicographic transducers.  They recognize relations on languages, unlike "plain" finite state automata (FSA) which recognize languages. If $M$ is some transducer, then we denote its recognized relation with $\mathcal{L}(M)$. Those relations are subsets of $\Sigma^*\times D$. The set of strings $\Sigma^*$ accepted by $M$ must be a regular language (indeed, if we erased output labels, we would as a result obtain FSA). The weights are erasable \cite{MendozaDrosik2020MultitapeAA}  in the sense that, give any lexicographic transducer we can always build an equivalent automaton without weighted transitions. If we didn't have $\tau$, the only output possible to be expressed for empty input would be an empty string as well. With $\tau$ we can express pairs like $(\epsilon,d)\in\mathcal{L}$ where $d\ne\epsilon$.  

The transducers can return at most finitely many outputs for any given input (see \textit{infinite superposition}\cite{MendozaDrosik2020MultitapeAA}). If we allowed for $\epsilon$-transitions (transitions that have $\epsilon$ as input label) we could build $\epsilon$-cycles and produce infinitely many outputs. However, automata that do so are not very interesting from practical point of view. Therefore we shall focus only on functional transducers, that is those which produce at most one output. If automaton does not have any $\epsilon$-cycles and is functional, then it's possible to erase all $\epsilon$-transitions (note that it would not be possible without $\tau$, because $\epsilon$-transitions allow for producing output given empty input). Therefore $\epsilon$-transitions don't increase power of functional transducers.

\begin{figure}[!t]
	\centering
	\includegraphics[width=9cm]{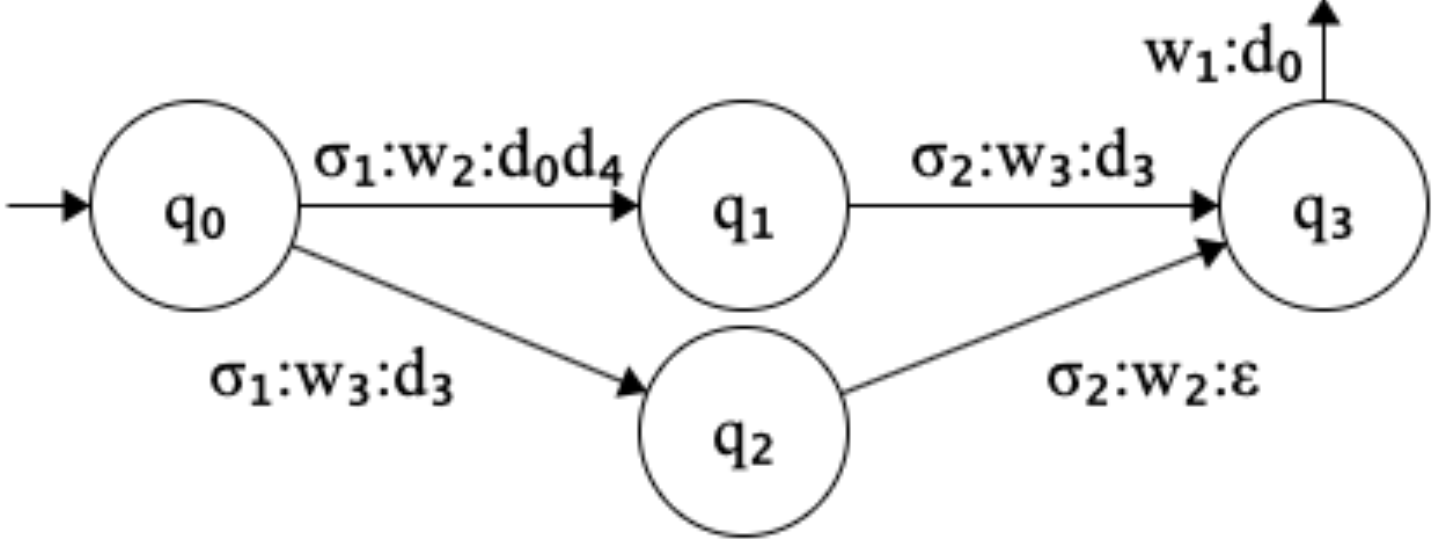} 
	\caption{Example of lexicographic transducer. State $q_0$ is initial. State $q_3$ in accepting, in the sense that $\tau(q_3)=(w_1,d_0)$. The remaining states have state output $\emptyset$.}
	\label{transducer}
\end{figure}

We say that transducer has \textbf{conflicting states} $q_1$ and $q_2$ if it's possible to reach both of them simultaneously (there are two possible routes with the same inputs and weights) given some input $\sigma$ and there is some another state $q_3$ to which both of those states can transition over the same input symbol $\sigma_i$. Alternatively, there might be no third state $q_3$, but instead both $q_1$ and $q_2$ have non-empty $\tau$ output (so $\tau$ can in a sense be treated like $q_3$). We say that transitions $(q_1,\sigma_i,w,d,q_3)$ and $(q_2,\sigma_i,w',d,q_3)$  are \textbf{weight-conflicting} if they have equal weights $w=w'$.  For instance in figure \ref{transducer} the states $q_1$ and $q_2$ are indeed conflicting because they both transition to $q_3$ over $\sigma_2$ but their transitions are not weight conflicting. It can be shown that transducers without weight-conflicting  transitions are functional. Moreover, if a transducer is functional but contains weight-conflicting transitions, then the weights can be reassigned in such a way that eliminates all conflicts \cite{MendozaDrosik2020MultitapeAA}. The only requirement is that there are enough symbols in $W$ (for instance, if $W$ had only one symbol, then all transitions of conflicting states would always be weight-conflicting). If there are at least as many weight symbols as there are states $\vert W \vert = \vert Q \vert$, then every functional transducer on $\vert W \vert$ states can be built without weight-conflicting transitions. For convenience we can assume that $W=\mathbb{N}$, but in practice all algorithms presented here will work with bounded $W$. Hence transducers without weight-conflicting transitions are equivalent in power to functional transducers. This is important because by searching for weight-conflicting transitions we can efficiently test whether transducer is functional or not.

\section{Ranged automata}

Often when implementing automata the algorithm behind $\delta$ function needs to efficiently find the right transition for a given $\sigma$ symbol. It's beneficial to optimise UNIX-style ranges like \texttt{[0-9]} or \texttt{[a-z]} as they arise often in practical settings. Even the \texttt{.} wildcard can be treated as one large range spanning entire $\Sigma$. If the alphabet is large (like ASCII or UNICODE), then checking every one of them in a loop is not feasible. A significant improvement can be made by only checking two inequalities like $\sigma_1\le x \le \sigma_{10}$, instead of large number of equalities. The current paper presents a way in which simplified model of $(\mathcal{S},k)$-automata\cite{MEER}\cite{Gandhi}, can be used to obtain major improvements. In particular we consider only automata that don't have any registers apart from constant values, that is $k=0$. Therefore we provide a more specialized definition of "ranged automata". 

Let $\Sigma$ be the (not necessarily finite) alphabet of automaton. Let $\chi$ be the set of subsets of $\Sigma$ that we will call \textbf{ranges} of $\Sigma$. Let  $\overline{\chi}$ be  the closure of $\chi$ under countable union and complementation (so it forms a sigma algebra). For instance, imagine that there is total order on $\Sigma$ and  $\chi$ is the set of all intervals in $\Sigma$. Now we want to build an automaton whose transitions are not labelled with symbols from $\Sigma$, but rather with ranges from $\chi$. Union $\chi_0\cup\chi_1$ of two elements from $\chi$ "semantically" corresponds to putting two edges, $(q,\chi_0,q')\in\delta$ (for a moment forget about outputs and weights) and $(q,\chi_1,q')\in\delta$. There is no limitation on the size of $\delta$. It might be countably infinite, hence it's natural that $\overline{\chi}$ should be closed under countable union. Therefore, $\chi$ is the set of allowed transition labels and $\overline{\chi}$ is the set of all possible "semantic" transitions. We could say that $\overline{\chi}$ is discrete if it contains every subset of $\Sigma$. An example of discrete $\overline{\chi}$ would be finite set $\Sigma$ with all UNIX-style ranges \texttt{[$\sigma$-$\sigma'$]} included in $\chi$. 

Another example would be set $\Sigma=\mathbb{R}$ with $\chi$ consisting of all ranges, whose ends are computable real numbers (real number $x$ is computable if the predicate $q<x$ is decidable for all rational numbers $q$). If we also restricted $\delta$ to be a finite set, then we could build effective automata that work with real numbers of arbitrary precision.

\section{Regular expressions}

Here we describe a flavour of regular expressions specifically extended to interplay with lexicographic transducers and ranged automata. 

Transducers with input $\Sigma^*$ and output $\Gamma^*$ can be seen as FSA working with single input $\Sigma^* \times \Gamma^*$. Therefore we can treat every pair of symbols $(\sigma,\gamma)$ as an atomic formula of regular expressions for transducers. We can use concatenation $(\sigma,\gamma_0)(\epsilon,\gamma_1)$ to represent $(\sigma,\gamma_0\gamma_1)$. It's possible to create ambiguous transducers with unions like  $(\epsilon,\gamma_0)+(\epsilon,\gamma_1)$.  To make notation easier, we will treat every $\sigma$ as $(\sigma,\epsilon)$ and every $\gamma$ as $(\epsilon,\gamma)$. Then instead of writing lengthy $(\sigma,\epsilon)(\epsilon,\gamma)$ we could introduce shortened notation $\sigma:\gamma$. Because we would like to avoid ambiguous transducers we can put restriction that the right side of $:$ should always be a string of $\Gamma^*$ and writing entire formulas (like $\sigma:\gamma_1+\gamma_2^*$) is not allowed. This restriction will later simplify Glushkov's algorithm. 

We define $\mathcal{A}^\Sigma$ to be the set of atomic characters. For instance we could choose $\mathcal{A}^\Sigma=\Sigma\cup\{\epsilon\}$ for FSA/transducers and $\mathcal{A}^\Sigma=\chi$ for ranged automata. 

We call  $RE^{\Sigma:D}$ the set of all regular expression formulas with underlying set of atomic characters $\mathcal{A}^\Sigma$ and allowed output strings $D$. It's possible that $D$ might be a singleton monoid $\{\epsilon \}$ but it should not be empty, because then no element would belong to $\Sigma^* \times D$. By inductive definition, if $\phi$ and $\psi$  are $RE^{ \Sigma:D}$  formulas and $d \in D$, then union $\phi + \psi$, concatenation $\phi \cdot \psi$, Kleene closure $\phi^*$ and output concatenation $\phi : d$ are $RE^{ \Sigma:D}$ formulas as well.  Define $V^{\Sigma:D}:RE^{\Sigma:D} \rightarrow \Sigma^* \times D$ to be the valuation function:  \\
$V^{\Sigma:D}(\phi + \psi) = V^{\Sigma:D}(\phi) \cup V^{\Sigma:D}(\psi)$ \\
$V^{\Sigma:D}(\phi \cdot \psi) = V^{\Sigma:D}(\phi) \cdot  V^{\Sigma:D}(\psi)$ \\
$V^{\Sigma:D}(\phi^*) = (\epsilon,\epsilon) + V^{\Sigma:D}(\phi) + V^{\Sigma:D}(\phi)^2 + ...$ \\
$V^{\Sigma:D}(\phi : d) = V^{\Sigma:D}(\phi)  \cdot (\epsilon,d)$ \\
$V^{\Sigma:D}(a) = a$ where $a\in\mathcal{A}^{\Sigma:D}$ \\
Some notable properties are: \\
$x:y_0 +x:y_1 = x:(y_0+y_1)$ \\
$x:\epsilon+x:y+x:y^2...=x:y^*$ \\
$(x:y_0)(\epsilon:y_1)  = x:(y_0y_1)$\\
$x_0:(y_0y')+x_1:(y_1y') = (x_0:y_0+x_1:y1)\cdot(\epsilon:y')$ \\
$x_0:(y'y_0)+x_1:(y'y_1) = (\epsilon:y')\cdot(x_0:y_0+x_1:y1)$  \\
Therefore we can see that expressive power with and without $:$ is the same. 

It's also possible to extend regular expressions with weights. Let $RE_W^{\Sigma: D}$ be a superset of $RE^{\Sigma: D}$ and $W$ be the set of weight symbols. If $\phi\in RE_W^{\Sigma\rightarrow D}$ and $w_0,w_1\in W$ then $w_0 \phi $ and  $\phi w_1 $ are in $RE_W^{\Sigma\rightarrow D}$. This allows for inserting weight at any place. For instance, the automaton from figure \ref{transducer} could be expressed using \[
((\sigma_1:d_0d_4)w_2(\sigma_2:d_3)w_3+
(\sigma_1:d_3)w_3\sigma_2w_2):d_0
\]
The definition of $V^{\Sigma:D}(\phi w)$ depends largely on $W$ but associativity $(\phi w_1) w_2 = \phi (w_1 + w_2)$ should be preserved, given that $W$ is a multiplicative monoid. This also implies that $w_1 \epsilon w_2 = w_1 w_2$, which is semantically equivalent to the addition $w_1 + w_2$. 

We showed that regular expressions for transducers can be expressed using pairs of symbols $(\sigma,\gamma)$. There is an alternative approach. We can encode both input and output string by interleaving their symbols like $\sigma_1\gamma_1\sigma_2\gamma_2$. Such regular expressions "recognize" relations rather than "generate" them. This approach has one significant problem. We have to keep track of the order. For instance, this $(\sigma_1\gamma_1\sigma_2 + \sigma_3)\gamma_4$ is a valid interleaved expression but this is not $(\sigma_1\gamma_1 + \sigma_3)\gamma_4$.

 In order to decide whether an interleaved regular expression is valid, we should annotate every symbol with its respective alphabet (like $(\sigma_1^\Sigma\gamma_1^\Gamma\sigma_2^\Sigma + \sigma_3^\Sigma)\gamma_4^\Gamma$). Then we rewrite the expression, treating alphabets themselves as the new symbols (for instance $(\Sigma\Gamma\Sigma + \Sigma)\Gamma$). If the language recognized by such expression is a subset of $(\Sigma\Gamma)^*$, then the interleaved expression valid. 

This leads us to introduce \textit{interleaved alphabets}. We should notice that $(\Sigma\Gamma)^*$ is in fact a local language. What it means is that in order to define interleaved alphabet we need 3 sets - set of initial alphabets $U$, set of allowed 2-factors of $V$ and set of final alphabets $W$. Moreover all the elements of $U$ must be pairwise disjoint alphabets. Similarly for $V$ if $(\Sigma_1,\Sigma_2)\in V$ and $(\Sigma_1,\Sigma_3)\in V$  then $\Sigma_2$ and $\Sigma_3$ must be disjoint. (For instance, in case of $(\Sigma\Gamma)^*$  we have $U=\{\Sigma\}$, $V=\{(\Sigma,\Gamma)\}$ and $W=\Gamma$). 

With interleaved alphabets we can encode much more complex "multitape automata". In fact it has certain resemblance to recursive algebraic data structures built from products (like $\{(\Sigma,\Gamma)\}$ in $V$) and coproducts (like $\{(\Sigma,\Gamma_1),(\Sigma,\Gamma_2)\}\in V$) .

It's possible to use interleaved alphabets together with $RE_W^{\Sigma:D}$ to express multitape inputs and mutitape outputs.

\section{Extended Glushkov's construction}

The core result of this paper is Glushkov's algorithm capable of producing very compact, $\epsilon$-free, weighted, ranged, functional, multitape transducers and automatically check if any regular expression is valid, when given specification of interleaved alphabets.

Let $\phi$ be some $RE_W^{\Sigma:D}$ formula. We will call $\Sigma$ the \textit{universal alphabet}. We also admit several subaphabets $\Sigma_1,\Sigma_2,...$ all of which are subsets of $\Sigma$. Each $\Sigma_i$ admits their own set of atomic characters $\mathcal{A}^{\Sigma_i}$ and we require that $\mathcal{A}^{\Sigma_i}\subset \mathcal{A}^{\Sigma}$.  Let $U_\Sigma,V_\Sigma,W_\Sigma$ be the interleaved alphabet consisting of all the subalphabets. For example $\Sigma$ could be the set of all 64-bit integers and then $V_\Sigma$ could contain its subsets like ASCII, UNICODE or binary alphabet $\{0,1\}$ (possibly with offsets to ensure disjointness). In cases when $D=\Gamma^*$, we can similarly define $U_\Gamma,V_\Gamma,W_\Gamma$, but there might be cases where $D$ is more a exotic set (like real numbers) and interleaved alphabet's don't make much sense. Moreover, we require $W$ to be a semiring. For instance, lexicographic weights have concatenation as multiplicative operation and $min$ is used for addition.

First step of Glushkov's algorithm is to create a new alphabet $\Omega$ in which every atomic character (including duplicates but excluding $\epsilon$) in $\phi$ is treated as a new individual character. As a result we should obtain new rewritten formula 
$\psi \in RE_W^{\Omega \rightarrow D} $ along with mapping $\alpha:\Omega \rightarrow\mathcal{A}^\Sigma$. This mapping will remember the original atomic character, before it was rewritten to unique symbol in $\Omega$.
For example 
\[
\phi=(\epsilon:x_0) x_0(x_0:x_1x_3)x_3 w_0+(x_1x_2)^* w_1
\]
will be rewritten as 
\[
\psi=(\epsilon:x_0) \omega_1(\omega_2:x_1x_3)\omega_3 w_0 + (\omega_4\omega_5)^* w_1
\]
with $\alpha= \{(\omega_1,x_0),(\omega_2,x_0),(\omega_3,x_3),(\omega_4,x_1),(\omega_5,x_2)\}$.

Every element $x$ of $\mathcal{A}^\Sigma$ may also be member of several subalphabets. For simplicity we can assume that all expressions are annotated and we know exactly which subalphabet a given $x$ belongs to. In practice, we would try to infer the annotation automatically and ask user to manually annotate symbols only when necessary.

Next step is to define function $\Lambda:RE_W^{\Omega\rightarrow D} \rightharpoonup ( D \times W)$. It returns the output produced for empty word $\epsilon$ (if any) and weight associated with it. (We use symbol $\rightharpoonup$ to highlight the fact that $\Lambda$ is a partial function and may fail for ambiguous transducers.) For instance in the previous example empty word can be matched and the returned output and weight is $(\epsilon,w_1)$. Because both $D$ and $W$ are monoids, we can treat $D \times W$ like a monoid defined as $(y_0,w_0)\cdot(y_1,w_1) = (y_0y_1,w_0+w_1)$. We also admit $\emptyset$ as multiplicative zero, which means that $(y_0,w_0)\cdot\emptyset=\emptyset$. We denote  $W$'s neutral element as $0$. This facilitates recursive definition: \\
$\Lambda(\psi_0+\psi_1) = \Lambda(\psi_0) \cup \Lambda(\psi_1)$ if at least one of the sides is $\emptyset$, otherwise error\\
$\Lambda(\psi_0\psi_1) =\Lambda(\psi_0) \cdot \Lambda(\psi_1)$ \\
$\Lambda(\psi_0 : y) = \Lambda(\psi_0) \cdot (y,0)$ \\
$\Lambda(\psi_0 w) = \Lambda(\psi_0) \cdot (\epsilon,w)$\\
$\Lambda(w \psi_0 ) =  \Lambda(\psi_0) \cdot (\epsilon,w)$ \\
$\Lambda(\psi_0^* ) = (\epsilon,0)$ if $(\epsilon,w) = \Lambda(\psi_0) $ or $\emptyset = \Lambda(\psi_0) $, otherwise error \\
$\Lambda(\epsilon) = (\epsilon,0)$\\
$\Lambda(\omega) = \epsilon$ where $\omega\in\Omega$

Next step is to define $B:RE_W^{\Omega\rightarrow D} \rightarrow (\Omega \rightharpoonup D \times W)$ which for a given formula $\psi$ returns set of $\Omega$ characters that can be found as the first in any string of $V^{\Omega\rightarrow D}(\psi)$ and to each such character we associate output produced "before" reaching it. For instance, in the previous example of $\psi$ there are two characters that can be found at the beginning: $\omega_1$ and $\omega_4$. Additionally, there is $\epsilon$ which prints output $x_0$ before reaching $\omega_1$. Therefore $(\omega_1,(x_0,0))$ and $(\omega_3,(\epsilon,0))$ are the result of $B(\psi)$. For better readability, we admit operation of multiplication $\cdot : (\Omega \rightharpoonup D \times W) \times (D \times W) \rightarrow (\Omega \rightharpoonup D \times W)$ that performs monoid multiplication on all $D \times W$ elements returned by $\Omega \rightharpoonup D \times W$. \\
$B(\psi_0 + \psi_1) = B(\psi_0)\cup B(\psi_1) $ \\
$B(\psi_0 \psi_1) = B(\psi_0) \cup \Lambda(\psi_0)\cdot B(\psi_1)$ \\
$B(\psi_0 w) = B(\psi_0)$ \\
$B(w \psi_0 ) = (\epsilon,w)\cdot B(\psi_0)$ \\
$B(\psi_0^*) =  B(\psi_0)$ \\
$B(\psi_0 : d) =  B(\psi_0)$ \\
$B(\epsilon) =  \emptyset$ \\
$B(\omega) =  \{(\omega,(\epsilon,0)) \}$ \\
It's worth noting that $B(\psi_0)\cup B(\psi_1)$ always yields function  (instead of relation) because every $\Omega$ character appears in $\psi$ only once and it cannot be both in $\psi_0$ and $\psi_1$. 

Next step is to define $E:RE_W^{\Omega\rightarrow D} \rightarrow (\Omega \rightharpoonup D \times W)$, which is very similar to $B$, except that $E$ collects characters found at the end of strings. In our example it would be $(\omega_3,(\epsilon,w_0))$ and $(\omega_5,(\epsilon,w_1))$. Recursive definition is as follows:\\ 
$E(\psi_0 + \psi_1) = E(\psi_0)\cup E(\psi_1) $ \\
$E(\psi_0 \psi_1) = E(\psi_0) \cdot \Lambda(\psi_1) \cup  B(\psi_1)$ \\
$E(\psi_0 w) = E(\psi_0) \cdot (\epsilon,w) $ \\
$E(w \psi_0 ) = E(\psi_0)$ \\
$E(\psi_0 ^*) =  E(\psi_0) $ \\
$E(\psi_0 : d) =  E(\psi_0) \cdot (d,0)$ \\
$E(\epsilon) =  \emptyset$ \\
$E(\omega) =  \{(\omega,(\epsilon,0)) \}$ 

Next step is to use $B$ and $E$ to determine all two-character substrings that can be encountered in $V^{\Omega\rightarrow D}(\psi)$. Given two functions $b,e:\Omega \rightharpoonup D \times W$ we define product $b \times e : \Omega \times \Omega \rightharpoonup  D \times W$ such that for any $(\omega_0,(y_0,w_0))\in b$ and $(\omega_1,(y_1,w_1)) \in c$ there is $((\omega_0,\omega_1),(y_0y_1,w_0+w_1)) \in b\times e$. Then define $L:RE_W^{\Omega\rightarrow D} \rightarrow (\Omega \times \Omega \rightharpoonup  D \times W)$ as: \\
$L(\psi_0 + \psi_1) = L(\psi_0)\cup L(\psi_1) $ \\
$L(\psi_0 \psi_1) = L(\psi_0) \cup  L(\psi_1) \cup E(\psi_0) \times B(\psi_1)$ \\
$L(\psi_0 w) = L(\psi_0) $ \\
$L(w \psi_0 ) = L(\psi_0)$ \\
$L(\psi_0 ^*) =  L(\psi_0) \cup E(\psi_0) \times B(\psi_0)$ \\
$L(\psi_0 : d) =  L(\psi_0)$ \\
$L(\epsilon) =  \emptyset$ \\
$L(\omega) =  \emptyset$  \\
One should notice that all the partial functions produced by $B$, $E$ and $L$ have finite domains, therefore they are effective objects from computational point of view. 

The last step is to use results of $L,B,E,\Lambda$ and $\alpha$ to produce automaton $(Q,q_\epsilon,W,\Sigma,D,\delta,\tau)$ with \\
$\delta:Q \times \Sigma \rightarrow (Q\rightharpoonup D\times W)$ \\
 $\tau:Q\rightharpoonup D \times W$ \\
$Q = \{q_\omega : \omega \in \Omega \} \cup \{q_\epsilon\}$ \\
$\tau = E(\psi)$ \\
$(q_{\omega_0},\alpha(\omega_1),q_{\omega_1},d,w)  \in \delta$ for every $(\omega_0,\omega_1,d,w) \in L(\psi) $ \\
$(q_\epsilon,\alpha(\omega),q_{\omega},d,w)  \in \delta$ for every $(\omega,d,w) \in B(\psi) $ 

This concludes the Glushkov's construction. Now it's possible to use specification $U_\Sigma$, $V_\Sigma$ and $W_\Sigma$ of interleaved alphabet to check if regular expression was correct. We can treat alphabets $\Sigma_1,\Sigma_2,...$ as colours and then colour each state with it's respective alphabet. If transition leads from state of colour $\Sigma_i$ to $\Sigma_j$ then we check that the pair $(\Sigma_i,\Sigma_j)$ is indeed present in $V_\Sigma$. Similarly we check colours of initial and accepting states.

\section{Optimisations}

The above construction can detect some obvious cases of ambiguous transducers, but it doesn't give us complete guarantee. We can check in quadratic time\cite{Marie-Pierre} for weight conflicting transitions to be sure. If there are none, then transducer must be functional. If we find at least one, it doesn't immediately imply that the transducer is ambiguous. In such cases we can warn the user and demand additional weight annotations in the regular expression. 

When $\mathcal{A}^\Sigma$ consists of all possible ranges  $\chi$, then the obtained $\delta$ is of the form $Q\times  W \times \chi \times D \times Q$. While, theoretically equivalent to $Q\times  W \times \Sigma \times D \times Q$, in practice it allows for more efficient implementations. For instance given two ranges \texttt{[1-50]} and \texttt{[20-80]}, we do not need to check equality for all $80$ numbers. The only points worth checking are $1,50,20,80$. Let's arrange them in some sorted array. Then given any number $x$, we can use binary search to find which of those points is closest to $x$ and then lookup the full list of intervals that $x$ is a member of. This approach works even for real numbers. More precise algorithm can be given a follows. Let $(x_0,y_0),(x_1,y_1),...(x_n,y_n)$ be closed ranges. Build an array \texttt{P} sorted in ascending order that contains all $y_i$ and also for every $x_i$ contains the largest element of $\Sigma$ smaller than $x_i$ (or more generally the supremum). Build a second array \texttt{R} that to every $i^{th}$ element of \texttt{R} assign list of ranges containing  $i^{th}$ element of \texttt{P}. Then in order to find ranges containing any $x$, run binary search on \texttt{P} that returns index of the largest element smaller or equal to $x$. Then lookup the list of ranges in \texttt{R}.

In Glushkov's construction epsilons are not rewritten to $\Omega$, which means that there are also no $\epsilon$-transitions. Hence we can use dynamic programming to efficiently evaluate automaton for any input string $x\in\Sigma^*$. The algorithm is as follows. Create two dimensional array $c_{i,j}$ of size $\vert Q \vert \times (\vert x \vert+1)$
where $i$-th column represents all nondeterministically reached states after reading first $i-1$ symbols. Each cell should hold information about the previously used transition. This also tells us the weight, output and source state of transition. For instance cell $c_{i,j}=k$ should encode transition coming from state $k$ to state $i$, after reading $j-1^{th}$ symbol. If state $q_i\in Q$ does not belong to $j^{th}$ superposition, then $c_{i,j}=\emptyset$. The first column is initialized with arbitrary value at $c_{i,1}=-1$ for $i$ referring to initial state $q_\epsilon$ and set to $c_{i,1}=\emptyset$ for all other $i$. Then algorithm progresses building next column from previous one. After filling out the entire array. The last column should be checked for any accepting states according to $\tau$. There might be many of them but the one with largest weight should be chosen. If we checked that the automaton has no weight-conflicting transitions, then there should always be only one maximal weight. Finally we can backtrack, to find out which path "won". This will determine what outputs need to be concatenated together to obtain path's output. This algorithm is quadratic $O(\vert Q\vert, \vert x \vert)$, but in practice each iteration itself is very efficient, especially when combined with binary search described in previous paragraph. By observing that states of automata are often sparsely connected, additional optimisation can be made by representing the two dimensional array with list of indices, as it's often done for sparse matrices.

%
%


\section{Conclusion}

Interleaved alphabets could find numerous applications with many possible extensions. In the context of natural language processing, they could be used to annotate human sentences with linguistic meta-information like parts of speech. Then transducers could built to take advantage of those tag. Using grammatical inference methods, one could also train such transducers to as POS taggers.

This approach cannot fully replace OpenFST, because it lacks their flexibility. The goal of OpenFST is to provide general and extensible implementation of many different transducer's, whereas the approach presented in this paper sacrifices extensibility for highly integrated design and optimal efficiency. For instance, Glushkov's algorithm could never support such operations  as inverses, projections, reverses or composition.


%

\appendices

\section*{Acknowledgment}

The authors would like to thank Piotr Radwan for all the inspiration.

\ifCLASSOPTIONcaptionsoff
  \newpage
\fi



%

%

\bibliographystyle{BibTeXtran}   
\bibliography{BibTeXrefs}       




\end{document}